\documentclass[fleqn,10pt]{wlpeerj}
\usepackage[T1]{fontenc}
\usepackage{hyperref}
\usepackage{multirow}
\usepackage{xcolor}
\newcommand{\noop}[1]{}

\title{Chilean Avian flu and its marine impacts: an online Statistical Process Control task}

\author[1]{Diego Carvalho do Nascimento}
\author[2,3]{Mauricio Ulloa}
\author[4]{Romulo Oses}
\author[5]{Francisco Louzada}
\author[5]{Oilson Alberto Gonzatto Junior}
\affil[1]{Departamento de Matemática, Facultad de Ingeniería, Universidad de Atacama, Copiapó, Chile}
\affil[2]{Veterinary Histology and Pathology, Institute of Animal Health and Food Safety, Veterinary School, University of las Palmas de Gran Canaria, las Palmas de Gran Canaria, Spain}
\affil[3]{Servicio Nacional de Pesca y Acuicultura (SERNAPESCA), Valparaíso, Chile}
\affil[4]{Centro Regional de Investigación y Desarrollo Sustentable de Atacama (CRIDESAT), Universidad de Atacama, Copiapó, Chile}
\affil[5]{Instituto de Ciências Matemáticas e de Computação (ICMC), Universidade de São Paulo, São Carlos, Brazil}
\corrauthor[1]{diego.nascimento@uda.cl}

\begin{abstract}
The rapid spread of the HPAI H5N1 virus, responsible for the Avian Flu, is causing a great catastrophe on the South American Pacific coast (especially in the south of Peru and north of Chile). Although very little attention has been delivered to this pandemic, it presents a tremendous lethal rate, though the number of infected humans is relatively low. Towards monitoring and statistical control, this work shows the Chilean national statistics from the year 2023, and presents the developed online tool for supporting the government's decision-making. Additionally, a Bayesian hierarchical spatiotemporal model was used to model the joint analysis of the weekly registered animal including spatial covariates as well as specific and shared spatial effects that take into account the potential autocorrelation between the HPAI H5N1 virus per Region. Our findings allow us to identify the hot-spot areas with high amounts of dead bodies (mostly pinnipeds and penguins) and their evolution over time.
\end{abstract}

\begin{document}

\flushbottom
\maketitle
\thispagestyle{empty}

\section{Introduction}
{ The South American Pacific coast boasts a remarkable and distinctive ecological ecosystem~\citep{arroyo1988effects}, largely shaped by its topography and the coastal influence of the Humboldt Current. Stretching along the western edge of South America, this powerful oceanic current brings cold, nutrient-rich waters up from the depths of the Pacific Ocean, creating a marine environment unlike any other on Earth~\citep{chavez2008northern}. This delicate balance has fostered the development of a rich and complex ecosystem including the microbial composition, for instance, see \cite{contreras2022chloroplast, Moskwa2020}. On the other hand, looking at the sea, a wide array of marine species thrives, i.e. massive whale flux, playful dolphins, and an abundance of seabirds \citep{norris2013marine}. The coastal cliffs of northern Chile provide essential nesting sites for numerous bird species, including the Humboldt penguin and marine mammals like the South American sea lion (\textit{Otaria flavescens}).}

In the opposite direction of animal life development, the emergence of animal pandemics (zooepidemic), the Highly pathogenic H5N1 avian influenza (HPAI H5N1) virus, presents a significant global health challenge \citep{kilpatrick2006predicting}. Unlike human diseases like COVID-19, this pandemic originates in animals and has complex transmissibility cycles that can extend from birds to marine and terrestrial species, complicating efforts to control and mitigate its impact \citep{chen2006establishment}. HPAI H5N1 primarily affected domestic flocks, wild migratory birds, and waterfowl like ducks and geese. These avian species can serve as reservoirs for the virus, which can then be transmitted to other animals, including marine species like pinnipeds and seabirds, through various mechanisms, such as contaminated water or direct contact.

Developing medications or vaccines for animals affected by these panzootic presents unique challenges~\citep{fan2009immunogenicity}. Unlike human vaccines, which can be administered relatively easily through injection or oral routes, delivering vaccines to wild marine animals is much more challenging~\citep{barnett2020ecological}. The sheer diversity and geographic dispersion of these animals make it difficult to reach them. Additionally, there may be ethical and environmental considerations when administering medications or vaccines to wildlife. Moreover, as these zooepidemics often involve a wide range of species, understanding the specific needs and susceptibilities of each one is a complex task. Research efforts are essential to developing effective interventions that not only protect animal populations but also reduce the risk of spillover into human populations, as some of these pathogens have the potential to mutate and pose significant threats to public health.

Recently, the evolution of the influenza A virus (IAV), detected in Chile, particularly concerning the AIV sub-type H5N1, which spread from Europe to North and Central/South America \citep{godoy2023evolution} highlights this threat. Chile's first record of highly pathogenic avian influenza (HPAI) was reported in December 2022. The occurrence of reassortant AIVs was demonstrated by the presence of genes from both North and South American isolates that have been found in wild birds since 2007. Later, molecular techniques and phylogenetic analysis detected the presence of HPAI A/H5N1 viruses, confirming valuable genomics insights into the Clade 2.3.4.4b in wild birds. \cite{jimenez2023detection} showed possible genetic mutations that did not correlate with known enhanced transmission or binding traits to mammalian receptors and, therefore, installing the urgency for detailed studies about potential risks of infection, transmission, and virulence. Remarkably, HPAI H5N1 has been detected in 2022 in wild birds \citep{jimenez2023detection}, sea lions \citep{ulloa2023mass}, and a human \citep{castillo2023first}. These findings emphasize the need for enhanced biosecurity on poultry farms and ongoing genomics surveillance to understand and manage AIVs in Chile's wild and domestic bird populations.

The research indicates that HPAVs, which are highly pathogenic avian influenza viruses, have spread widely among wild birds in Europe, Asia, and Africa \citep{sagong2022emergence}. In the past, the H5N1 virus spread from Europe to North and Central/South America. Scientists have confirmed that all the circulating HPAI H5N1 viruses originated from influenza A/goose/Guangdong/1/1996, which only affected birds \citep{gilbertson2023mammalian}. It has been demonstrated that reassortant AIVs, avian influenza viruses, have genes from both North and South American isolates and have been found in wild birds since 2007. Bioinformatic comparative analysis of gene sequences confirmed valuable genomics insights into Clade 2.3.4.4b, evidencing the presence of HPAI A/H5N1 viruses in wild birds, demonstrating that Clade 2.3.4.4 viruses originated from clade 2.3.4 and increasing capacity to reassort, giving rise to H5Nx viruses with novel NA pairings. Gilbertson and his colleagues have found that the evolution of this lineage gave rise to Clade 2.3.4.4b viruses, where the H5 re-acquired an N1 NA pairing. These viruses threaten public health and food security worldwide due to their zoonotic potential and poultry economic losses. \cite{chmielewski2011avian} and his team have also confirmed this fact in their research published in 2011.

Generally, outbreaks of avian influenza occur sporadically during autumn and decrease by spring. However, since 2022, there have been continued outbreaks during the summer in Europe and North America, as reported by \cite{harvey2023changing}. The HPAI H5N1 clade 2.3.4.4b was first detected in North America in early 2022, most likely introduced by migratory wild birds from Europe, as stated by \cite{gunther2022iceland}. Starting in late 2022, the virus spread throughout South America, as reported by \cite{leguia2023highly}.

Chile recorded its first HPAI in December 2022, which was mainly caused by the AIV sub-type H5N1. The infection was found in a Pelecanus thagus specimen, as reported by \cite{ariyama2023highly}. Since then, there have been reports of AIV H5N1 infections in various wild birds, as documented by \cite{jimenez2023detection} and \cite{CDC.2023}. Additionally, marine mammals have also been affected, which was reported by \cite{ulloa2023mass} and \cite{SAG.2023}. A human case was reported by \cite{castillo2023first}, and \cite{SERNAPESCA.2023}.

A recent study by \cite{godoy2023evolution} examined the evolution of different subtypes of IAV in birds and mammals, including humans, in Chile. The study focused particularly on the current status of HPAI H5N1 viruses. The findings suggest that the distribution and spread of AIV H5N1 in Chile is a result of a complex interplay between ecological and human factors. The study found that there is a negative correlation between the distance to the closest urban center, precipitation, and temperature seasonality, which suggests a potential for the virus to cross over to Antarctica/subantarctic islands. These viruses' presence in Chile highlights the need for increased biosecurity measures on poultry farms and continuous genomic surveillance approaches to understand and control AIVs in both wild and domestic bird populations in Chile.

Considering the big challenge to unravel patterns related to the zooepidemic dissemination, Spatio-temporal models using Integrated Nested Laplace Approximation (INLA) represent a cutting-edge approach to geospatial modeling and data analysis \citep{lindgren2015bayesian}. INLA is a versatile and efficient estimation framework that allows researchers to explore and understand complex relationships in data with both spatial and temporal components and computational low-cost~\citep{gomez2020bayesian, blangiardo2015spatial}. These models have found applications in a wide range of fields, from epidemiology and environmental science to social sciences and beyond \citep{pavani2023joint, adin2023alleviating}.

The integration of statistical control mechanisms combined with the Spatio-temporal INLA model \citep{blangiardo2013spatial} enhances the artificial intelligence monitoring models, which play a pivotal role in enabling government intervention and decision-making based on inferential/predictive insights. By continuously collecting and analyzing data from various sources, AI-driven forecasting models can detect emerging trends, anomalies, and potential risks with a high degree of accuracy. These insights empower government agencies to proactively respond to critical issues, whether by predicting disease outbreaks, monitoring environmental changes, or anticipating economic shifts. By harnessing the power of real-time data and predictive analytics, governments can make informed policy decisions, allocate resources efficiently, and take preemptive measures to address societal challenges, ultimately leading to more effective governance and improved public welfare.

{ This study tested/adjusted a spatiotemporal statistical modeling approach to analyze and monitor the dynamic of the 2023 first semester (weekly records) of the Chilean coastal border report by the Fisheries and Aquaculture Service of Chile (SERNAPESCA) of protected species (marine mammals, sea turtles and penguins). This methodology enables quantifying the spatial covariates and shared spatial effects across the Chilean Regions, taking into account the potential autocorrelation between the HPAI H5N1 virus impact. As a result, a Shiny framework compiles statistical descriptive and spatiotemporal control charts for six groups of animals decomposed per Chilean region (expected pattern versus observed).}

\section{Material \& Methods}\label{sec:MM}

The Chilean government is sampling animals (mostly wild birds and marine species) reported as dead, physically emaciated, or disoriented that may probably be infected by Highly Pathogenic Avian Influenza Virus H5N1 (HPAIv H5N1). 
A great number of carcasses, on the coastal border, have been reported during 2023 by the official data set of the Fisheries and Aquaculture Service of Chile (SERNAPESCA). Towards the end of the South American winter season a reduction in the wild migratory birds volume, and marine birds nesting was observed according to the Agricultural and Livestock Service of Chile (SAG) data set which is the institution responsible for monitory and guarding Chilean's wild and domestics birds except penguins.

This study was based on the SERNAPESCA database, which covers the analysis from January to June (first semester) of 2023. This database reported 17,691 registered marine animals including marine mammals, sea turtles and penguins. This database presents 20 variables/features (shown in Table \ref{tab:features}), which are related to the location of the observed animal and its characteristics, and further information about the collection characterization.

\begin{table}[htbp]
\centering
\caption{Features present in the SERNAPESCA dataset.}
\begin{tabular}{ccc}
\hline
REGION & RECORD (n) & Latitude (LAT) \& Longitude (LON) \\
Sample TIME & SPECIES Type & INSTITUTIONS ENROLLED \\
GENDER & MARKS & REHABILITATION CENTER \\
AGE & CITY & VITAL CONDITION \\
SIZE & H5N1 SAMPLED & LOCATION INFORs \\
STARTING DAY & ENDING DAY & CORPORAL CONDITION \\ \hline
\end{tabular}
\label{tab:features}
\end{table}

This study proposed an online framework to help the Chilean government with pattern extraction and to link the HPAI H5N1 pandemic evolution to society, as an informative source. To achieve this goal, it was developed a data-driven methodology and deployed on the web through the combination of the open-source R software with the Shiny app. The Chilean collected carcasses (exclusively marine mammals, sea turtles, and penguins) online information is hosted at \url{https://cemeai.shinyapps.io/H5N1/}, supported by SERNAPESCA. The next subsections will discuss, in detail, the Shiny app structure, as well as the Artificial Intelligence model adopted in this study. Brief information about the HPAI H5N1 genomics structure and virus evolution can be seen at the Appendix \ref{APX}.

\subsection{The Shiny app}

The Shiny app is a user-friendly and versatile tool that encapsulates the open-source script from R programming \citep{r2023}, offering an innovative platform for the deployment of statistical and machine-learning models~\citep{shiny2023}. Shiny's hallmark is its accessibility, making it a valuable resource for individuals with no level of programming to do data analysis without much effort. With Shiny, users can create interactive web applications without having to delve deep into the complexities of AI programming scripts. This simplicity empowers researchers, and analysts to make decisions in real-time with a broader audience, making data-driven more accessible to stakeholders in various fields.
\begin{figure}[!ht]
    \centering
    \includegraphics[width=\textwidth]{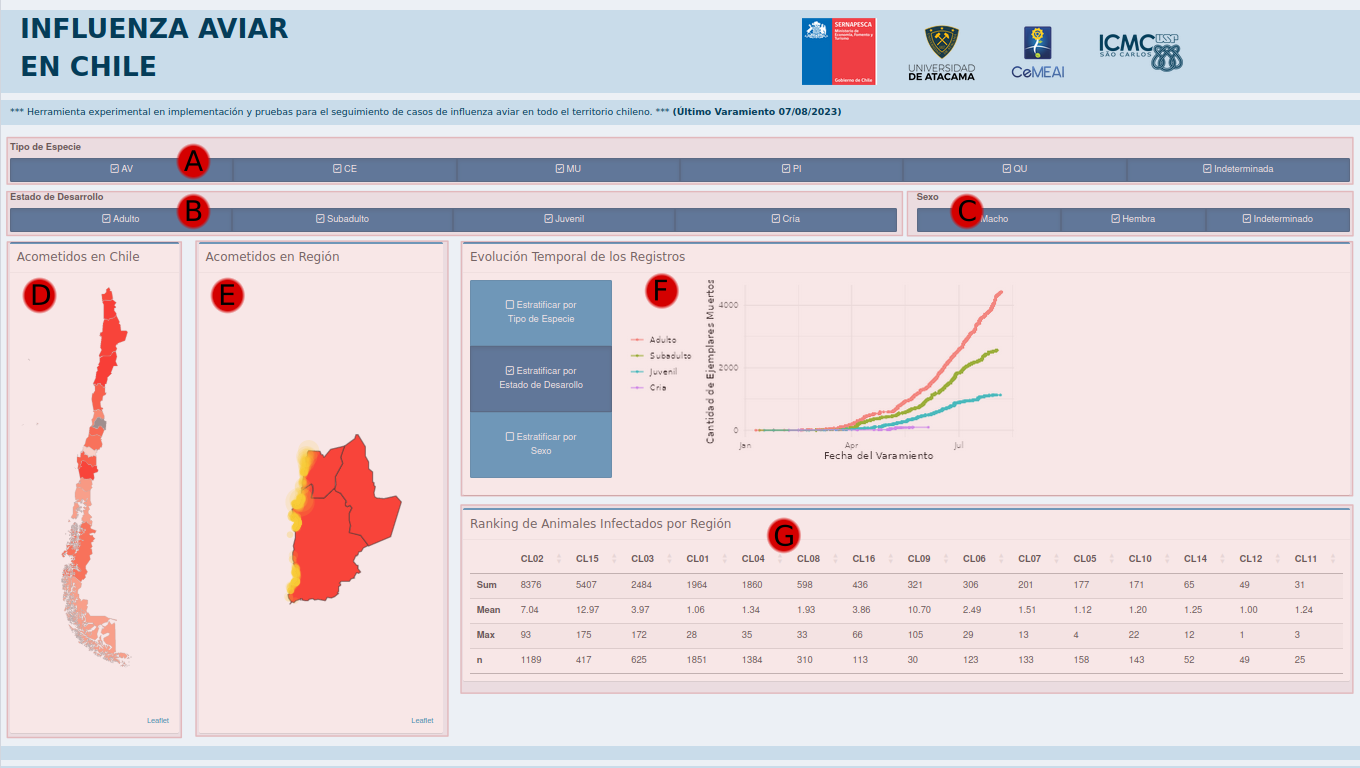}
    \caption{Overview of the developed framework. Seven elements are highlighted: A) Specie type, B) Animal's aging, C) Animal's gender, D) Number of registered animal distributed across Chile, where the red color is the highest rate of dead animals and white is the lowest density rate, E) Selected state/region, F) Cumulative plot, which can be represented grouping by Species, Aging or Gender, and G) The rank per region of i) number of total registered animal, Sum; ii) mean of registered animal, Mean, iii) maximum registered animal, Max, and iv) the number of field visits, n.}
    \label{fig:FRAME1}
\end{figure}

One of the most compelling aspects of the Shiny app is its ability to integrate statistical and machine-learning models seamlessly. This article proposed an integration of statistical tools wrapped in a user-friendly interface, allowing non-technical users to generate real-time insights regarding SERNAPESCA's registered animals. This democratization of advanced analytics is particularly valuable in government and industry reports, where data-driven can lead to significant improvements in informative outcomes.

\begin{figure}
    \centering
    \includegraphics[width=\textwidth]{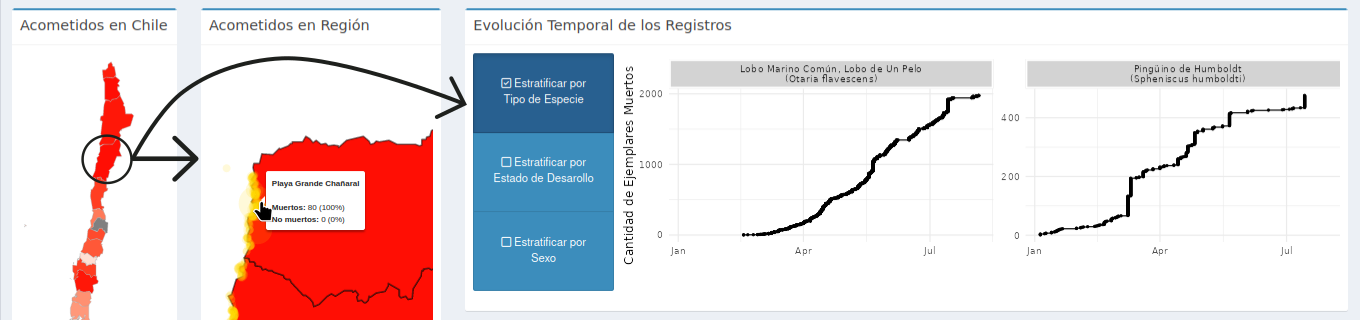}
    \caption{Framework central part enables the selection of the Chilean's region (left-hand); then the next plot (center-frame) enables the zoom-in and -out highlighting the registered animal given its georeference (LAT, LON), as well as the summarization of each record such as the number of dead animals vs. not dead. Finally, the right-hand shows the empirical cumulative frequency given the selected filter (grouping by species, aging, or gender). }
    \label{fig:FRAME2}
\end{figure}

Figure \ref{fig:FRAME1} shows the complete overview of the developed framework, which highlights the seven interactive elements that can be useful based on the summarization of the required detail. For instance, the user can filter per Region or Species type the latest dynamic information of the observed cumulative records. Figure \ref{fig:FRAME2} shows the central part of the proposed framework that accommodates the filter for selecting the cumulative dynamic of the desired region (if no region is selected then it will be analyzed the whole Chilean country), which can be also desegregated per Species, Aging or Gender.

\begin{figure}[!ht]
    \centering
    \includegraphics[width=\textwidth]{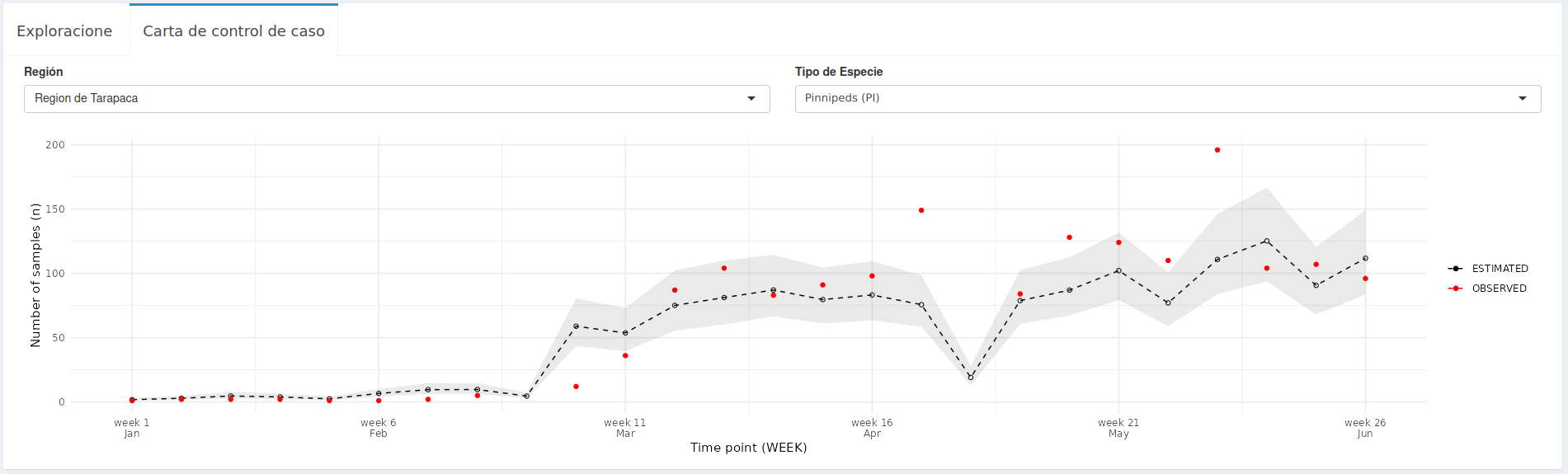}
    \caption{Statistical Control Chart monitoring based on the proposed spatio-temporal model, i.e. Pinnipeds (PI) in the Region of Tarapacá. The top tabs indicate the SPC monitoring tabs (\textit{Carta de control de caso}), as well as the selected options for the REGION and ANIMAL TYPE. Then, the dashed-line and shade bound are associated with the adjusted model (expected pattern), and the red-dots are the observation samples. Moreover, the red dots above the expected pattern (gray shade) highlight an anomalous/high-trend.}
    \label{fig:SPC_Spatio-temporal}
\end{figure}

In order to assess the statistical monitoring of the registered animal count, a spatio-temporal model was adopted to model the expected number per Region considering each Species. For that, a Statistical Process Control (SPC) control chart was implemented for countable response based on the Spatial joint model, explained in detail in the next subsection. Figure \ref{fig:SPC_Spatio-temporal} shows an illustration of the output of the weekly dynamic per region by animal species, whereas the dashed-line and shade bound are associated with the adjusted model (expected pattern based on the adjusted model), and the red-dots are the observations. It can be said that every red dot higher than the expected bound is an out-control week observation (associated with an increase of observed carcasses locally).

The adopted spatio-temporal AI model helps to understand the knowledge-based dynamic of the Chilean coastal dead animals, moreover to unravels the zooepidemic trend and cycle (spatial and temporal estimate). In this manner, it was adopted a Bayesian hierarchical spatial model for the joint analysis (spatial covariates and shared spatial effects) by decomposing the explicability (process variation), which one can understand the associated features with the death of animals whereas geographic patterns to identify risk factors. Particularly, this class of statistical learning models is useful in delivering Explainable artificial intelligence frameworks.

\subsection{Spatio-temporal AI model}

\cite{louzada2021spatial} presented an overview of the most common spatial correlation inclusion between latent effects, under the Bayesian approach. In which, most of them include a prior distribution with a spatial relationship. Such as the Conditional Autoregressive (CAR), Besag–York–Mollie (BYM), Leroux CAR, and Stochastic Partial Differential Equation (SPDE), whereas may (or not) consider the spatial distribution as a Gaussian Markov Random Fields (GMRF). This spatial term is summed to a mean process decomposition, in the regression form, regardless of the adopted spatial relationship/explanation (covariance estimation).

For instance, let's consider a countable response variable $Y^{(s)}_{i,t}$ the observed number of cases per $s$ grouping, at point area $i$, and on time $t$ respectively. Most often, the literature countable response is modeled as a Poisson process~\citep{Nelder1972, best2000spatial}, nevertheless, whenever the supposition of the mean process is not equal to its variance especially when a zero inflation case occurs (called overdispersion), it might not be appropriated~\citep{Lambert1992, Hall2000}. Alternatively, the Tweedie distribution is defined by $\theta:=\{ \mu, p, \sigma^2 \}$ parameters, where $\mu \in \mathbb{R}$ is the mean, $\sigma^2 > 0$ is the dispersion and $p \in ( -\infty,0] \cup [1, \infty )$ the power parameters~\citep{tweedie1984index, jorgensen1997theory}. 
In this manner, this work adopted a hierarchical formulation for the spatial model as follows:
\begin{align}
    Y^{(s)}_{i,t} | \mu_{i,t},p,\sigma^{2(s)}_{i,t} &\sim \text{TWEEDIE}(\mu_{i,t},p,\sigma^{2(s)}_{i,t})\\
    \mathbb{E}[Y^{(s)}_{i,t}]&=\mu_{i,t}=g^{-1}(X_i^T \pmb{\beta} + Z_{i,t}^T \pmb{\gamma}),
\end{align}

\noindent where the process relationship can be translated as a cross-sectional dataset, ($Y^{(s)}_{i,t}, X, Z_{i,t}$), where $i = \{1, 2, \dots, 5.841\}$ observations, $t = \{1, 2, \dots, 26\}$ week, $s = \{1, 2, \dots, 6\}$ month. The $Y$'s are the weekly growth of the number of animals collected/analyzed in the first semester of 2023, $X$ are the feature design matrix which includes Region, Species Type and their interaction, and $Z$ is the spatio-temporal latent effects matrix (Latitude \& Longitude, and time format in Week). The $\pmb{\beta}$  is a vector of unknown regression coefficients, $\pmb{\gamma}$ is a vector of unknown latent effects coefficients and $g(\cdot)$ is the log link function whenever $p=2$ for the Tweedie family. 

Then, based on the collected data, the regression structure was
\begin{align}
    log (\mu_{i,t}) =& \beta_0 + \beta_1 \text{Species}_{PI} + \cdots + \beta_5 \text{Species}_{MU} + \beta_6 \text{Region}_{\#2} + \cdots + \beta_{19} \text{Region}_{\#16} + \\ 
    &\beta_{20} I(\text{Region}_{\#2} \times \text{Species}_{PI}) + \beta_{21} I(\text{Region}_{\#2} \times \text{Species}_{CE}) +\\
    & \cdots  + \beta_{89} I(\text{Region}_{\#16} \times \text{Species}_{MU}) +  Z(s, t)\pmb{\gamma} + \xi(s, t)
\end{align}

\noindent where essentially the theoretical model adopted only two coraviables, and their interaction, here represented by I. Additionally, it is important to mention that each feature turned its classes into binary variables. Considering the covariables \textit{Animals Species Type} (BI, PI, CE, MU, QU, UND) and \textit{Regions} (\#1, ..., \#16, not existing the \#13). Reminding that one class, of each variable, is used as the model intercept. Therefore, the intercept, $\beta_0$, is a mixed effect of the categories BIRDS animal species (BI, which are only penguins) in the TARAPACÁ Region (\#1). Finally, the $Z(s, t)$ is a set of covariates observed in the location s and period t, and $\xi(s, t)$ is the spatial random effects represented by the Gaussian process continuously projected in space and with a covariance function of the Matérn class.

The model to be fit is on a temporal dependence considering an autoregressive, AR(1), on the week time-period, as well as an AR(1) for the spatial dependence, and random walk, RW(1) for the spatio-temporal fit to the monthly spatial distribution. The resulting latent effect is a GMRF with zero mean and covariance matrix presented by \cite{simpson2016going}, where it originated from a product of two terms, $\sigma^{2(s)}_{i,t} = \tau \Sigma_b \bigotimes \Sigma_w$, where $\tau$ is a weighting factor, $\Sigma_b$ is the structure of the covariance between the different groups (the spatio-time format in Months per species) and $\Sigma_w$ is the structure of the covariance matrix within each group (the spatio-time format in Weeks of each species). Then, the INLA formula structure adopted was

\begin{verbatim}
formula = y ~ 1 + Region + Type + Type*Region +
  f(spatial.field, model = spde, group = spatial.field.group, 
    control.group = list(model = "ar1") +
  f(ID.Week, model = "ar1") + f(Region, Month, model = "rw1")
\end{verbatim}

One key advantage of spatio-temporal models with INLA is their ability to capture intricate patterns, regardless of the great amount of parameters to be estimated, including dependencies over space and time while providing fast and accurate estimates. INLA employs a combination of Bayesian techniques and numerical approximations to handle complex spatial and temporal structures in a computationally efficient manner. This allows researchers to extract valuable insights from large datasets and address real-world problems where both spatial and temporal dimensions play a crucial role. This study considered the spatial grid as a Stochastic Partial Differential Equation (SPDE) process, modeling the amount of registered animals over time and geography or assessing the impact of changes as a continuous mesh, valuable tools for decision-makers and researchers. 

Furthermore, these models provide an intuitive platform for uncertainty quantification, enabling users to understand the inherent variability in their data and make more informed decisions. INLA's probabilistic framework offers a natural way to incorporate uncertainties in both parameter estimates and model predictions, which is particularly important in applications where risk assessment and decision-making are critical. As the field of spatio-temporal modeling continues to evolve, INLA plays a pivotal role in simplifying the development of sophisticated models that can accommodate both spatial and temporal dimensions, ultimately contributing to more accurate predictions and informed policies in a variety of domains.

\section{Results} \label{sec:res}

Chile's fauna \& flora have been considered unique and relatively closed ecosystems since the Andes Alps blocked the entry of plagues and other non-local species from the Atlantic Ocean main-land countries. Additionally, to give a hint about the importance of the Northern Chilean fauna \& flora preservation, 33.19\% (41k out of 125k) of the sea lion population is located there, and between 10-12 thousand adults Humboldt Penguins nesting in Pan de Azúcar island, Isla Grande \& Chañaral (Atacama region), Isla Choros (Coquimbo region) and Isla Cachagua (Valparaíso region).

\begin{figure}[h!]
    \centering
    \includegraphics[width=\textwidth]{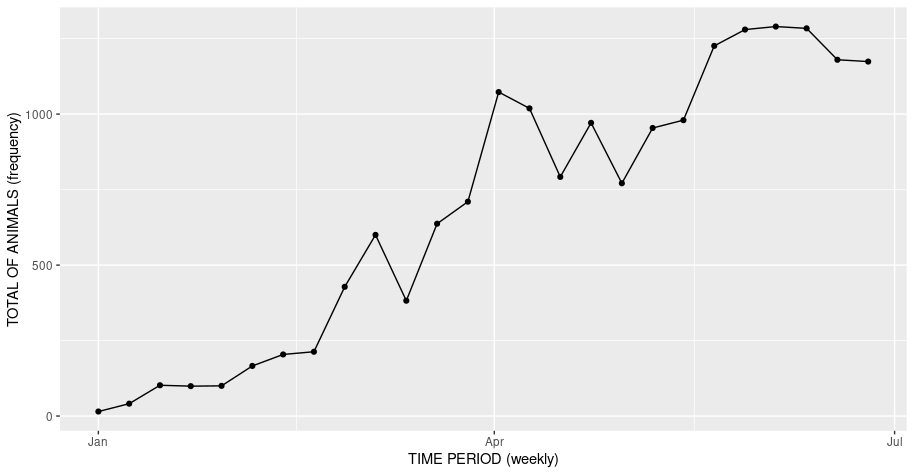}
    \caption{Weekly increase of the number of animals collected/analyzed in the first semester of 2023.}
    \label{fig:weekly}
\end{figure}

Moreover, the north of Chile is known to be a great transitivity region for many animals of the South American Pacific coast, such as penguins, marine mammals, and, other wild birds due to its huge productivity. Most of this phenomenon is impulsed by the Atacama Trench. 
However, all this is at risk given the HPAI H5N1 zooepidemic~\citep{ulloa2023mass}. In the first semester of 2023, based on SERNAPESCA's records (exclusively marine mammals, sea turtles, and penguins), it was analyzed a total of 17,691 animals. Figure \ref{fig:weekly} shows the weekly dynamic of the number of animals collected/analyzed, mostly carcasses. The expected harmonic monthly increase was 164\% (that is, the average monthly odds ratio), whereas the biggest gain was from February to March. 

Table \ref{tab:State} shows the dynamic of registered animals across the Chilean coastal regions. It is important to mention that the observations were obtained based on the community reports. Whereas the higher volume originated from the Antofagasta Region (\#2), Arica \& Parinacota (\#15), and Atacama (\#3), all three are located in the North of Chile. Though, based on average the higher was Arica \& Parinacota Region (\#15), Araucanía (\#9) [not related to HPAI H5N1], and Antofagasta (\#2) based on the number of trips to the field vs collected data. { The mean is associated with the average carcasses collected per week (based on the announced appearance on the coast), as greater the region was highlighted in red, therefore the map is describing the regions more infected.}
\begin{table}[htbp]
\caption{SERNAPESCA registration of carcasses ranked per Region (`23 first semester).}
\includegraphics[width=\textwidth]{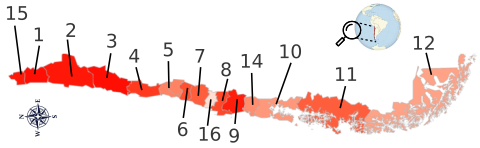}
\resizebox{\textwidth}{!}{\begin{tabular}{l|ccccccccccccccc}
\hline
REGION \# & 2 & 15 & 3 & 1 & 4 & 8 & 16 & 9 & 6 & 7 & 10 & 5 & 14 & 11 & 12 \\ \hline
TOTAL & 5,494 & 4,603 & 2,014 & 1,887 & 1,460 & 580 & 425 & 318 & 279 & 192 & 164 & 159 & 62 & 29 & 25 \\ 
MEAN (weekly) & 6.04 & 13.5 & 3.69 & 1.04 & 1.21 & 1.99 & 4.17 & 11.4 & 2.74 & 1.54 & 1.21 & 1.10 & 1.27 & 1.26 & 1 \\ \hline
\end{tabular}}
\label{tab:State}
\end{table}

Beholding the spatio-temporal descriptive, a greater concern is towards the animal species which is been affected, { suggesting a higher weekly mean prevalence in the North of Chile area}. Since Chile's fauna and flora are so unique, catastrophic damage can be unleashed if not well monitored. Table \ref{tab:group} presents some descriptive statistics about the impacts on the different groups of animals. {The weekly evolution regardless of the class is not greater than one carcass, nevertheless, some time-points were highly unexpected. These maximum values reinforce the need for a temporal Statistical Process Control analysis.} For instance, the Pinnipeds (PI) and Birds (BI, which are only penguins considered) are the species groups mostly seen as deteriorating in the first semester of 2023.
\begin{table}[htbp]
\centering
\caption{ {Descriptive of the weekly number of animals' carcasses registered per marine classification (considering records from the '23 first semester).} Despite this, Chelonians have not been affected by the HPAI H5N1 virus.}
\begin{tabular}{l|ccccccc}
\hline
 Animal Class & \multicolumn{1}{l}{   Min.} & \multicolumn{1}{l}{ 1st Qu.} & \multicolumn{1}{l}{  Median} & \multicolumn{1}{l}{    Mean} & \multicolumn{1}{l}{ 3rd Qu.} & \multicolumn{1}{l}{    Max.} & \multicolumn{1}{l}{Total} \\ \hline
Pinnipeds (PI) & 1 & 1 & 1 & 3 & 1 & 175 & 14,887 \\ 
Birds (BI) & 1 & 1 & 1 & 3 & 1 & 105 & 2,646 \\ 
Cetaceans (CE) & 1 & 1 & 1 & 1 & 1 & 4 & 91 \\ 
Mustelids (MU) & 1 & 1 & 1 & 1 & 1 & 1 & 35 \\ 
Chelonians (QU) & 1 & 1 & 1 & 1 & 1 & 4 & 24 \\
Undefined (UND) & 1 & 1 & 1 & 1 & 1 & 2 & 8 \\  \hline
\end{tabular}
\label{tab:group}
\end{table}

Table \ref{tab:species} highlights, per species, from the total of 17,691 sampled animals. For instance, the South American sea lion's carcasses collected correspond to 83.9\% of the records, which is related to 15\% of the estimated population in Chile (19k out of 130k). Not less importantly, the statistics associated with the Humboldt penguin correspond to 30\% of the Chilean's population.
\begin{table}[htbp]
\centering
\caption{Total Species ranked/recorded  ('23 first semester). Grey-scale are species carcasses collected for analysis, though verified not to be related to the HPAI H5N1 virus.}
\begin{tabular}{rc}
\hline
\multicolumn{1}{c}{SPECIES} & \multicolumn{1}{l}{TOTAL} \\ \hline
Lobo Marino Común, Lobo de Un Pelo - Otaria flavescens & 14,840 \\ 
Pingüino de Humboldt - Spheniscus humboldti & 2,224 \\ 
Pingüino de Magallanes - Spheniscus magellanicus & 421 \\ 
Chungungo - Lontra felina & 34 \\ 
Lobo Fino de Juan Fernández - Arctocephalus philippii & 24 \\ 
Marsopa Espinosa - Phocoena spinipinnis & 22 \\ 
{\color{gray} Tortuga Verde - Chelonia mydas} & 20 \\ 
Delfín Chileno - Cephalorhynchus eutropia & 16 \\ 
Ballena Sei, Rorcual Bacalao, Rorcual de Rudolphi - Balaenoptera borealis & 15 \\ 
Elefante Marino - Mirounga leonina & 12 \\ 
Lobo Fino Austral - Arctocephalus australis & 10 \\ 
Delfín Naríz de Botella - Tursiops truncatus & 9 \\ 
Delfín Oscuro - Lagenorhynchus obscurus & 8 \\ 
Undefined & 8 \\ 
Ballena Jorobada - Megaptera novaeangliae & 4 \\ 
Delfín Común - Delphinus delphis & 4 \\ 
{\color{gray} Tortuga Olivácea - Lepidochelys olivacea} & 4 \\ 
Cachalote - Physeter macrocephalus & 2 \\ 
Delfin Gris - Grampus griseus & 2 \\ 
Orca - Orcinus orca & 2 \\ 
Ballena de Aleta, Rorcual Común, Fin - Balaenoptera physalus & 1 \\ 
Ballena Franca Austral - Eubalaena australis & 1 \\ 
Ballena Picuda de Cuvier - Ziphius cavirostris & 1 \\ 
Cachalote Enano de Cabeza Corta, Cachalote Pigmeo - Kogia breviceps & 1 \\ 
Delfín Austral - Lagenorhynchus australis & 1 \\ 
Delfín de Diente Áspero - Steno bredanensis & 1 \\ 
Foca Leopardo - Hydrurga leptonyx & 1 \\ 
Huillín - Lontra provocax & 1 \\ 
Pingüino Rey - Aptenodytes patagonicus & 1 \\ 
Zifio de Arnoux - Berardius arnuxii & 1 \\ \hline
\end{tabular}
\label{tab:species}
\end{table}

Once knowing that the two species most affected are the South American sea lion (\textit{Otaria flavescens}) and the Humboldt Penguin (\textit{Spheniscus humboldti}), Table \ref{tab:StateSpecies} summarized the top 10 locations. The North of Chile represents greater alert (Regions 1, 2, 3, 4, and 15) than the South of Chile (Regions 8 and 16). 
\begin{table}[htbp]
\centering
\caption{The top 10 recorded species per Region ('23 first semester) are from the Pinnipeds and Birds/Penguins animal species.}
\begin{tabular}{lcr}
\hline
\multicolumn{1}{c}{REGION (\#)} & SPECIE & \multicolumn{1}{c}{TOTAL} \\ \hline
Antofagasta (2) &  Lobo Marino Común, Lobo de Un Pelo - Otaria flavescens & 5,234 \\ 
Arica (15) &  Lobo Marino Común, Lobo de Un Pelo - Otaria flavescens & 3,617 \\ 
Tarapacá (1) &  Lobo Marino Común, Lobo de Un Pelo - Otaria flavescens & 1,811 \\ 
Atacama (3) &  Lobo Marino Común, Lobo de Un Pelo - Otaria flavescens & 1,572 \\ 
Coquimbo (4) &  Lobo Marino Común, Lobo de Un Pelo - Otaria flavescens & 944 \\ 
Arica (15) &  Pingüino de Humboldt - Spheniscus humboldti & 943 \\ 
Bío-Bío (8) &  Lobo Marino Común, Lobo de Un Pelo - Otaria flavescens & 541 \\ 
Coquimbo (4) &  Pingüino de Humboldt - Spheniscus humboldti & 485 \\ 
Atacama (3) &  Pingüino de Humboldt - Spheniscus humboldti & 430 \\ 
Ñuble (16) &  Lobo Marino Común, Lobo de Un Pelo - Otaria flavescens & 411 \\ \hline
\end{tabular}
\label{tab:StateSpecies}
\end{table}

In order to extend the discussion about the descriptive evidence, this study adopted a Bayesian spatio-temporal model to infer possible trends and extend generalizations. The analysis considered credible interval was 80\%. Based on the fixed effects (in Table \ref{tab:model}), the Pinnipeds and Cetaceans animal type presented statistical significance compared to the Birds category (only penguins considered). Whereas, the regions of Arica, Antofagasta, Atacama, Araucanía, Los Lagos, Los Ríos, and Ñuble presented statistical significance considering Tarapacá as a baseline.

\begin{table}[htbp]
\centering
\caption{Fixed Effects posterior estimations from the spatio-temporal model. Every Region feature (factor) is compared with the Region \#1 (Tarapacá).}
\begin{tabular}{rl|cccccc}
\hline
& & \multicolumn{1}{c}{Mean} & \multicolumn{1}{c}{SD} & \multicolumn{1}{c}{ Q10} & \multicolumn{1}{c}{ Q50} & \multicolumn{1}{c}{Q90} & \multicolumn{1}{c}{Mode} \\ \hline
\multicolumn{2}{r|}{Pinnipeds vs BIRDS} & -0.425 & 0.116 & -0.574 & -0.425 & -0.277 & -0.425\\
\multicolumn{2}{r|}{Cetaceans vs BIRDS} & -0.451 & 0.276 & -0.805 & -0.451 & -0.097 & -0.451\\ 
\multicolumn{2}{r|}{Region \#2 (Antofagasta)} & -0.917 & 0.581 & -1.667 & -0.914 & -0.172 & -0.91 \\
\multicolumn{2}{r|}{Region \#3 (Atacama)} & 1.121 & 0.733 & 0.179 & 1.125 & 2.059 & 1.125\\ 
\multicolumn{2}{r|}{Region \#9 (Araucanía)} & 3.498 & 1.131 & 2.058 & 3.485 & 4.956 & 3.483\\ 
\multicolumn{2}{r|}{Region \#10 (Los Lagos)} & 2.903 & 1.096 & 1.532 & 2.857 & 4.339 & 2.689\\ 
\multicolumn{2}{r|}{Region \#14 (Los Ríos)} & 2.287 & 1.101 & 0.905 & 2.247 & 3.725 & 2.245\\ 
\multicolumn{2}{r|}{Region \#15 (Arica)} & 2.525 & 0.797 & 1.501 & 2.527 & 3.547 & 2.527\\ 
\multicolumn{2}{r|}{Region \#16 (Ñuble)} & 1.882 & 0.921 & 0.714 & 1.869 & 3.068 & 1.869\\ \hline
PI-Region \#2  & \parbox[t]{1mm}{\multirow{14}{*}{\rotatebox[origin=c]{90}{vs BIRDS - Region \#1 (Tarapacá)}}} & 1.584 & 0.135 & 1.41 & 1.584 & 1.757 & 1.584  \\
PI-Region \#3 & & 0.361 & 0.149 & 0.17 & 0.361 & 0.551 & 0.361 \\
PI-Region \#4 & & 0.306 & 0.124 & 0.147 & 0.306 & 0.466 & 0.306 \\ 
PI-Region \#5 & & 0.407 & 0.185 & 0.17 & 0.407 & 0.644 & 0.407 \\ 
PI-Region \#6 & & 0.913 & 0.244 & 0.601 & 0.913 & 1.226 & 0.913 \\ 
PI-Region \#7 & & 0.749 & 0.332 & 0.324 & 0.749 & 1.175 & 0.749 \\ 
PI-Region \#8 & & 0.855 & 0.207 & 0.589 & 0.855 & 1.12 & 0.855 \\
PI-Region \#9 & & -1.925 & 0.327 & -2.344 & -1.926 & -1.506 & -1.926 \\ 
PI-Region \#16 & & 1.873 & 0.288 & 1.504 & 1.873 & 2.243 & 1.873 \\
CE-Region \#3 & & -1.052 & 0.483 & -1.671 & -1.052 & -0.432 & -1.052 \\ 
CE-Region \#15 & & -2.421 & 0.344 & -2.862 & -2.421 & -1.981 & -2.421 \\ 
CE-Region \#16 & & 1.738 & 0.486 & 1.116 & 1.738 & 2.361 & 1.738 \\ 
MU-Region \#3 & & -0.795 & 0.493 & -1.427 & -0.795 & -0.163 & -0.795 \\ 
MU-Region \#15 & & -2.282 & 0.472 & -2.888 & -2.282 & -1.676 & -2.282 \\ \hline
\end{tabular}
\label{tab:model}
\end{table}

Furthermore, regions \#9 and \#14 (Araucanía and Los Rios) showed zero dead animals by Avian Flu (based on HPAI H5N1 sample tests). Then, all carcasses were related to bycatch and not to the HPAI virus. The reason why they are shown to be statistically significant is that in comparison to the Tarapacá region (\#1 as baseline), their performance was statistically different. 

Table \ref{tab:model} also shows the interaction between the features \textit{Species} and \text{Region location} observing a few of the heterogeneity explainability across the space. A greater difference in the \textit{pinnipeds} pattern between the Tarapacá region v/s (\#2, 3, 4, 5, 6, 7, 8, 9, and 16), since this category includes the South American sea lion and Humboldt Penguin. Nevertheless, the North of Chile (in the Regions of Atacama, \#3, and Arica \& Parinacota, \#15) showed prevalence towards the Cetaceans \& Mustelids category against the Birds (Penguins only) - Tarapacá Region. An interesting fact is that the Region of Ñuble (\#16), located in the South of Chile, showed alert based on the adjusted model, especially in the Pinnipeds and Cetaceans category saw/registered as disoriented animals, but not high cases of dead ones. This is important to mention that the Region of Ñuble stranded animals were live ones that were returned to the sea. 

Furthermore, since the explainable AI model has a linear structure all regions are necessarily included to estimate these spatial effects/contributions. For instance, explaining the estimation based on the obtained parameters, let's consider just the deterministic part (fixed effects, that is, regardless of the spatial-temporal explanation) comparing the data from Araucanía v/s Arica regions (\#9 and \#15). That is, based on the national (Chilean) pattern and some regional contribution, the expected Pinnipeds result would be like,
\begin{align}
    \mathbb{E}[Y_{i}|\text{Pinnipeds in Region \#9}] &= \text{exp(} \widehat{\text{Intercept}} + \widehat{\text{Pinnipeds}} + \widehat{\text{Region \#9}} + \\
    & \ \ \ \ \ \ \widehat{\text{Interaction} }\text{[Pinnipeds $\times$ Region \#9])}\\
    &=exp(-0.246-0.425+3.498-1.925) \approx 2
\end{align}

\noindent though the point-estimation is not zero (adopting the posterior means), the interval estimation includes it (so statistically this expectation is equal to zero). Nevertheless, numerically the Araucanía shows three times less than Arica \& Parinacota region (\#15) Pinnipeds expectation,
\begin{align}
    \mathbb{E}[Y_{i}|\text{Pinnipeds in Region \#15}] &= \text{exp(} \widehat{\text{Intercept}} + \widehat{\text{Pinnipeds}} + \widehat{\text{Region \#15}} + \\
    & \ \ \ \ \ \ \widehat{\text{Interaction} }\text{[Pinnipeds $\times$ Region \#15])}\\
    &=exp(-0.246-0.425+2.525+0.010) \approx 6
\end{align}

\noindent reminding that the intercept is the mixed effect of the categories of animal species BIRDS (only penguins considered) within the TARAPACÁ REGION (\#1), therefore, all estimated parameters are an increase (or decrease) from this baseline. Nevertheless, the greatest explanations are related to the temporal and spatial effects (random/latent), discussed in the following.

Table \ref{tab:RandEffect} shows the posterior distributions for each adopted latent effect (spatial, temporal, and spatio-temporal) and the two other Tweedie parameters (power parameter, p, and dispersion, $\sigma^2$).

\begin{table}[htbp]
\centering
\caption{Random/Latent effects posterior estimations from the spatio-temporal model.}
\begin{tabular}{l|cccccc}
\hline
 & \multicolumn{1}{l}{Mean} & \multicolumn{1}{l}{SD} & \multicolumn{1}{l}{Q10} & \multicolumn{1}{l}{Q50} & \multicolumn{1}{l}{ Q90} & \multicolumn{1}{l}{Mode} \\ \hline
p parameter for Tweedie & 1.92 & 0.004 & 1.914 & 1.92 & 1.925 & 1.92 \\ 
Dispersion parameter for Tweedie & 0.437 & 0.008 & 0.427 & 0.437 & 0.447 & 0.437 \\ 
Range for spatial.field & 1.732 & 0.467 & 1.181 & 1.678 & 2.35 & 1.578 \\ 
Stdev for spatial.field & 1.568 & 0.218 & 1.302 & 1.552 & 1.857 & 1.516 \\
GroupRho for spatial.field & 0.907 & 0.034 & 0.862 & 0.912 & 0.945 & 0.921 \\
Precision for ID.Week & 0.116 & 0.037 & 0.074 & 0.111 & 0.164 & 0.101 \\ 
Rho for ID.Week & 0.999 & 0 & 0.999 & 0.999 & 1 & 1 \\
Precision for Region & 39.302 & 32.514 & 12.084 & 30.264 & 76.64 & 18.11 \\ \hline
\end{tabular}
\label{tab:RandEffect}
\end{table}

Additionally, the posterior mean estimation of the power parameter ($\widehat{p}$) was 1.92 with $CI_{80\%}$[1.914, 1.925], and dispersion ($\widehat{\sigma^2}$) was 0.437  with $CI_{80\%}$[0.427, 0.447]. That combination leads the Tweedie distribution shape towards the (Geometric) Compound-Poisson-gamma probability distribution \citep{kokonendji2021tweedie}. 

\begin{figure}
    \centering
    \includegraphics[width=\textwidth]{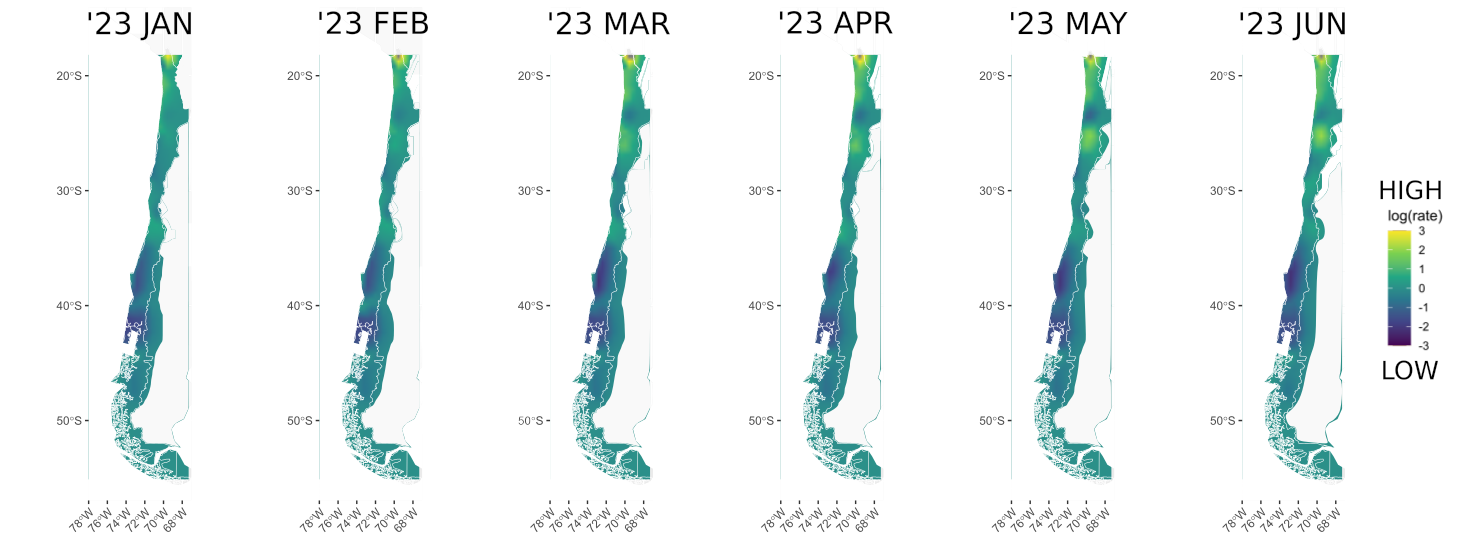}
    \caption{Time-varying spatial effect of the weekly number of animals collected/analyzed in the first semester of 2023.}
    \label{fig:spatial}
\end{figure}

Visually, the spatial dynamic estimation across the first semester of 2023 is presented in Figure \ref{fig:spatial}. The first highlight point is that the region of Arica \& Parinacota (\#15), latitude 20ºS (top region), always showed the highest concentration of field-visiting registered animals. The region of Atacama (\#3), latitude 27ºS, is visually showing a rapid increase in observed animals. Finally, no region from the South of Chile is highlighted, latitude 35ºS or greater, that is they did not showed a high density of records (not at the beginning nor the end of the observed period of the '23 first semester).

\begin{figure}
    \centering
    \includegraphics[width=\textwidth]{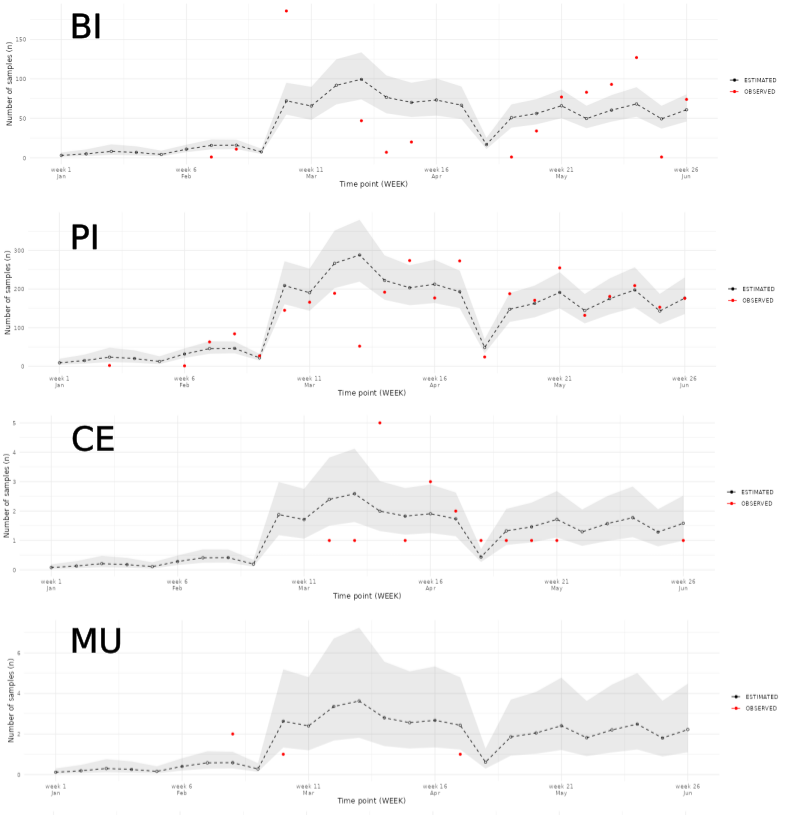}
    \caption{Weekly dynamic of the Arica \& Parinacota Region (\#15) per animals type (’23 first semester).}
    \label{fig:State15-SPC}
\end{figure}

For instance, Figure \ref{fig:State15-SPC} shows the estimated dynamic originating from the proposed spatio-temporal model for the Region of Arica \& Parinacota (\#15) per animal type. The observed points beyond the estimated bounds show outlier patterns which can be seen as higher than expected (based on the trade-off between the National and Regional hierarchical structure). To obtain more estimates from other regions, use the Shiny application developed/presented in this work.

All findings are in the same direction as published by \cite{ulloa2023mass}, which detailed the mortality of the South American sea lions (\textit{Otaria flavescens}) in Chile in the first semester of 2023. The infection and its transmission are verified through the digestive tract given the sharing of the feeding habitat between wild birds carriers of the HPAI H5N1 and marine mammals and Humboldt penguins. Interestingly, 94\% of the mortality in marine mammals, especially common sea lions, was registered within the geographical limits of the Northern macrozone of Chile (From Arica to Coquimbo), where the Chilean fisheries of Sardines and Anchovies operate making these group of species feed upon the same school of fish (i.e. small fishes) and making the virus available in the sea environment. This scenario did not occur in southern Chile (which presented an HPAI H5N1 mortality of less than 6\%) since the southern sea lion population normally feeds upon other prey, such as hakes, salmon, cod, etc., species of larger fishes and is unattractive to wild birds, therefore no sharing of feeding habitat occurs.

\begin{figure}
    \centering
    \includegraphics[width=\textwidth]{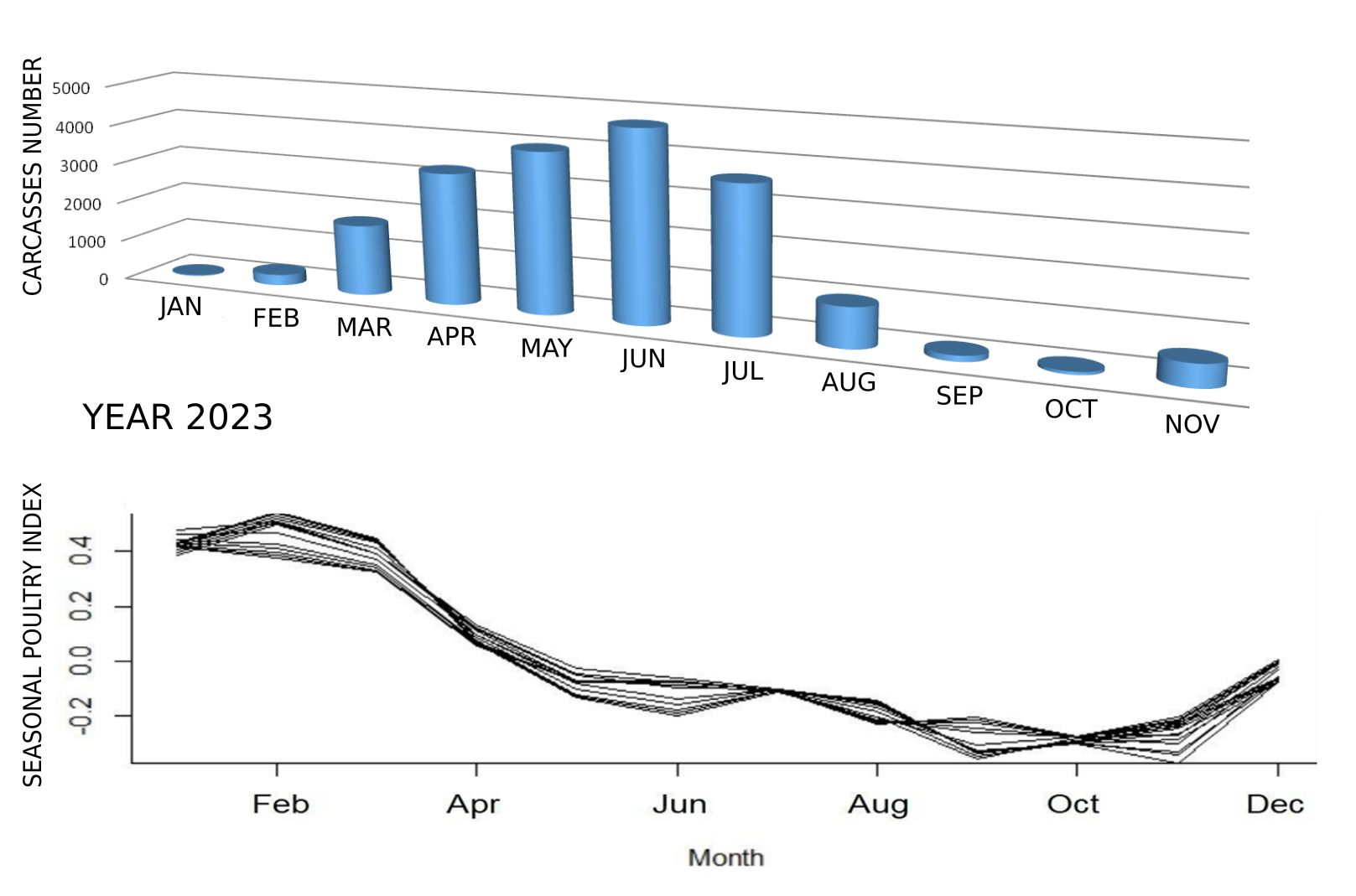}
    \caption{Illustrates the temporal cycle shifting association between HPAI H5N1 virus at common sea lions vs poultry. In the top chart, the SERNAPESCA records HPAI H5N1 prevalence in common sea lions, and, in the bottom chart, the WAHIS estimation of the seasonal incidence of HPAI H5N1 in poultry.}
    \label{fig:TREND}
\end{figure}

Additionally, in terms of predictions for next season, we must bear in mind that the incidence of HPAI H5N1 in poultry occurs with high numbers of cases during the months of December, January, and February starting to decrease during March and April to maintain low incidence during the rest of the year. Conversely, the only known prevalence of marine mammal deaths starts to rise in March maintaining high mortality rates until August (the H5N1-related hypothesis can be extended til September or October). Figure \ref{fig:TREND} illustrates this temporal cycle shifting supposition, in the top chart, the SERNAPESCA records of HPAI H5N1 prevalence in common sea lions, and, the bottom chart, World Animal Health Information System (WAHIS) estimation of the seasonal incidence of HPAI H5N1 in poultry.

On top of all, climate change is inducing transformative effects on marine ecosystems, and one notable consequence is the emergence of new microorganisms (or mutations) displaced by oceanic currents. In the Atacama Ecosystem, which is intricately linked to the dynamics of the Humboldt Current~\citep{Thiel2007Humboldt}, the impact of these newly introduced microorganisms is particularly pronounced. The delicate balance of the Atacama Ecosystem, known for its unique and adapted flora and fauna, is eminently to be disrupted as these microorganisms (e.g. HPAI H5N1) interact with existing species and alter ecological relationships. This phenomenon underscores climate change's intricate and cascading effects on marine ecosystems, emphasizing the need for surveillance strategy and comprehensive studies (such as this one) to understand and mitigate the repercussions on biodiversity and ecosystem functioning.

\section{Discussion}\label{sec:con}
Chile's northern fauna and flora are considered unique heritage due to its status as a biogeographic island, which means that it is naturally isolated by the Atacama Desert in the north and by the Andes Mountains. This means that many species of flora and fauna, are only found in Chile and nowhere else on the planet, known as endemism~\citep{morrone2018evolutionary}. {Moreover, the Andean Alps act as a natural barrier to Chile from species and plagues, therefore, useful for highlighting a new entry of invasive organisms and estimating their dynamism whenever something happens.} A big concern in the Chilean scientific community is related to the not much government support for removing the carcasses (faster and properly), increasing the risk of the virus acquisition by wild and especially domesticated mammals (dogs and cats, which live freely part of the daylight on the streets).

This study thoroughly addresses a spatiotemporal methodology employed in the data carcasses collection, overcoming the lack of positive antigen testing and autopsies to confirm the presence of the HPAI H5N1, though presenting as future research gaps bolstering the trustworthiness of its results in epidemiological inquiries. A notable deficiency in government support, in the second semester of 2023, for carcass removal raises (and registration) concerns regarding potential biases in data collection, particularly in disease surveillance. Additionally, the mid-2023 marine surveillance data for HPAI H5N1, indicates a gap/limitation in data coverage, demanding new solutions to overcome this and increase the robustness of the spatiotemporal presented model applied to forecasting the expected amount of carcasses for 2024.

Additionally, \cite{rzymski14avian} discussed the potential of A/H5N1 to adapt better to mammalian hosts, including humans. Additionally, given their geographical location which combines the Pacific Ocean with Humboldt \& the Andes Alps, Chile's north coast and Peruvian south coast are a potential surveillance strategy to detect and mitigate the threat posed by these highly pathogenic viruses. Their status as a biogeographic island is useful for the monitoring task e.g. avian influenza, as a reference location and its dynamism is useful as a proxy to South America (or even worldwide).

It's worth noting that the adopted methodology employed by the Bayesian hierarchical spatiotemporal model, separates global patterns from regional ones, thus also considering spatial and time dynamics together in its estimation process. This model is particularly useful in addressing individual or isolated observations arising from data registration limitations, as it breaks down the global-regional contribution into a hierarchy. Improving data quality would require collaboration between the public sector (government), academia, and citizens to ensure timely data registration across Chile's vast geographic expanse. Lastly, integrating this methodology with the proposed Shiny framework could be an effective online statistical process-control tool for monitoring and decision-making support.

Regardless of the mathematical complexity of spatio-temporal models, the Shiny app can serve as an effective online statistical process-controlling tool for monitoring and decision-making support. Organizations can monitor and analyze data streams, identifying deviations from established norms or quality thresholds. When anomalies or trends that require immediate attention are detected, alerts can be generated within the Shiny app, enabling swift corrective actions. This online monitoring capability is invaluable in manufacturing or surveillance, where it can help prevent defects and optimize production, as well as in healthcare or epidemiology, where it can aid e.g. pandemic dynamism. Overall, the Shiny app's user-friendly interface and integration with R's analytical capabilities make it a powerful tool for data-driven decision-making and even real-time process control across various industries.

Finally, spatiotemporal effects were statistically detected highlighting the importance of monitoring the HPAI H5N1 dynamic in Chile (especially in the northern area). These elements are not only related to the avian flu virus but also new highly pathogenic viruses speed up by the weather change. 

Although, knowing that periodic data collection is implemented on Chile's coastal animal monitoring, SERNAPESCA had the SAG support on the marine surveillance (towards HPAI H5N1 dynamic) only in the first semester of 2023. Therefore, it will be a limitation for further modeling to be carried out without a data correction strategy (beyond June 2023). Additionally, every registered/collected animal body was submitted to an antigen test, although many resulted in a negative HPAI H5N1 prevalence. Nonetheless, newer findings suggest the accuracy of HPAI H5N1 detection is through autopsy in the animal's brain or lung \citep{ulloa2023mass}. Despite this, is undeniable that the amount of animal carcasses, on the Chilean coast, has exponentially increased over 2023, requiring further and deeper microbiological studies of its contents.


\section*{Conflict of Interest Statement}
The authors declare that the research was conducted in the absence of any commercial or financial relationships that could be construed as a potential conflict of interest.

\section*{Author Contributions}

All authors listed have made a substantial, direct, and intellectual contribution to the work and approved it for publication.

\section*{Funding}
All authors are grateful for the undirected funding partially provided by the SERNAPESCA, SAG, and Universidad De Atacama. Francisco Louzada acknowledges support from the São Paulo State Research Foundation (FAPESP grant no. 2013/07375-0) and CNPq (grant no. 301976/2017-1).

\section*{Acknowledgments}
We thank the SERNAPESCA and the SAG personnel for their support and contributions, especially in the sample collection and diagnostic. All authors also acknowledge the contributions of Iván Miranda Lértora, Claudio Ramírez De La Torre, and Soledad Tapia (all from SERNAPESCA-Chile), which aided the efforts of this study's authorizations and information support.

\section*{Data Availability Statement}
The datasets analyzed in this study can be found in the \url{https://github.com/ProfNascimento/H5N1}.

\bibliography{ref.bib}

\appendix
\section{APPENDIX - The HPAI H5N1 review}\label{APX}

Influenza viruses belong to the Orthomyxoviridae family \citep{Koonin2019}, order Articulavirales \citep{schoch2020ncbi}, which involve four genera: Alphainfluenzavirus, Betainfluenzavirus, Gammainfluenzavirus, and Deltainfluenzavirus. The Orthomyxoviridae family has a segmented negative-sense single-stranded RNA genome with 6-8 incorporated segments. Each genus has a single ratified species: influenza A virus (IAV), influenza B virus (IBV), influenza C virus (ICV), and influenza D virus (IDV), respectively \citep{parry2020divergent}. IAV exhibits the highest genetic variability and the broadest host range among the four types of influenza viruses. According to the type of viral envelope glycoproteins, it is possible to classify two IAVs: rod-shaped haemagglutinin (HA) and mushroom-shaped neuraminidase (NA). The current record indicates that 16 HA types and nine NA types of IAV have been described \citep{wille2020ecology}, and all of these subtypes, which have a worldwide distribution, are considered to have originated from the silent reservoir in the wild aquatic birds \citep{fouchier2005characterization}.

Influenza viruses infect many wild birds and mammals, including humans, adapting to new hosts and forming several lineages specific to humans, horses, swine, dogs, and other animals \citep{yang2021bat}. The IAVs can be classified as avian influenza viruses (AIVs), swine influenza viruses (SIVs), or other animal influenza viruses, depending on their origin host. In human populations, seasonal epidemics and occasional pandemics are caused by mutations in the genomes of human influenza viruses.

The IAV subtypes that circulate in bird populations can mutate into the highly pathogenic avian influenza (HPAI) phenotype, which causes severe disease in poultry and high mortality rates. These mutations frequently occur in the AIV H5 and H7 subtypes \citep{uribe2022molecular}. Later, another variation the H9N2 AIV was reported in human infections highlighting potential epidemiological characteristics~\citep{tan11reported}. Conversely, viruses that cause mild diseases in poultry are referred to as low pathogenic avian influenza (LPAI) viruses. 

Since 2022, outbreaks of HPAI H5N1 viruses have been detected in Europe, then North America \citep{kessler2021influenza}, and in Latin America in commercial poultry, backyard poultry, wild birds, and mammals, including humans \citep{olsen2006global}. The ongoing circulation of the virus in all global regions, except Oceania, remains a significant concern, encompassing both poultry and other animal species \citep{kandeil2023rapid}. In early December 2022, an increase in mortality was observed among wild birds, primarily pelicans, along the north coast of Chile. SAG collected samples from domestic and wild birds to detect HPAI and conduct an epidemiological investigation. Among the species affected by AIV, pelicans (Pelecanus thagus) were the most frequently infected, accounting for 54\% of the cases. They were followed by vultures (Cathartes aura) and Peruvian boobies (Sula variegata), which indicate their susceptibility to AIV infection.

All the sequenced samples were identified as belonging to the H5N1 subtype and the 2.3.4.4b H5 clade. Comparisons with sequences available on the GISAID database revealed that A/Peru/LIM-003/2022 and A/Peru/LAM-002/2022 (GISAID Isolate IDs EPI\_ISL\_16249730 and EPI\_ISL\_16249681) exhibited the closest genetic similarity in terms of the HA gene. Similar results were also observed when analyzing the phylogenies of the NA and internal genes, genomics further supporting the genetic relatedness of these samples to previously identified strains. The analysis supports the introduction of the HPAI H5N1 clade 2.3.4.4b into the Americas via the Atlantic Flyway. Furthermore, it suggests that the virus has spread to other migratory bird routes within the region. This finding indicates the potential for significant dissemination of the HPAI H5N1 clade 2.3.4.4b strain among bird populations across different migratory pathways in the Americas \citep{SAG.2023}.

Recently, the evolution of different IAV subtypes in birds and mammals, including humans, in Chile, with an emphasis on the current status of HPAI H5N1 viruses, was reviewed \cite{godoy2023evolution}. This work concluded that the distribution and spread of AIV H5N1 in Chile are the product of a complex interplay between ecological and human factors. The hypothesis founded relies on a negative correlation with the distance to the closest urban center and precipitation and temperature seasonality, suggesting a potential to cross over to Antarctica/subantarctic islands. This suggests that the presence of these viruses in Chile underscores the need for increased biosecurity on poultry farms and continuous genomic surveillance approaches to understanding and controlling AIVs in both wild and domestic bird populations in Chile. This evidence highlights the great need to develop online tools for constant monitoring to support governmental data-driven politics.

\end{document}